\documentclass[sigconf]{acmart}

\usepackage{soul}
\usepackage{makecell}
\usepackage{balance}
\usepackage{framed}

\AtBeginDocument{%
  }

\usepackage{ifthen}
\usepackage{xcolor}
\usepackage[normalem]{ulem}

\provideboolean{revising}
\setboolean{revising}{false}

\ifthenelse{\boolean{revising}}{
    \newcommand{\revision}[1]{\textcolor{blue}{#1}}

    \newcommand{\delete}[1]{\textcolor{red}{\sout{#1}}}

    \newcommand{\sgcomment}[1]{%
        \textcolor{purple}{\textbf{[SG: #1]}}%
    }
    \newcommand{\tlcomment}[1]{%
        \textcolor{teal}{\textbf{[TL: #1]}}%
    }

    \newcommand{\tc}[1]{%
        \textcolor{orange}{\textbf{[TODO: #1]}}%
    }
    
}{
    \newcommand{\revision}[1]{#1}
    \newcommand{\delete}[1]{}
    
    \newcommand{\sgcomment}[1]{}
    \newcommand{\tlcomment}[1]{}
    
    \newcommand{\tc}[1]{}
    
}

\copyrightyear{2026}
\acmYear{2026}
\setcopyright{cc}
\setcctype{by}
\acmConference[UIST '26]{The 39th Annual ACM Symposium on User Interface Software and Technology}{November 02--05, 2026}{Detroit, MI, USA}
\acmBooktitle{The 39th Annual ACM Symposium on User Interface Software and Technology (UIST '26), November 02--05, 2026, Detroit, MI, USA}
\acmDOI{10.1145/3830398.3830543}
\acmISBN{979-8-4007-2856-3/2026/11}

\newcommand{\toolname}{\textsc{\textbf{BITE}}}

\newenvironment{mybox}[1]{%
  \par\medskip
  \noindent
  \colorbox{blue!15}{%
    \parbox{\dimexpr\linewidth-2\fboxsep\relax}{\textbf{#1}}%
  }%
  \par\noindent
  \MakeFramed{\advance\hsize-\width \FrameRestore}%
}{%
  \endMakeFramed
  \medskip
}

\begin{document}

% \title{BITE: Privacy Preserving \textit{B}rowsing-\textit{I}nformed \textit{T}racing of AI \textit{E}ngagement }

% \title{Towards Longitudinal Human–LLM Alignment through Contextual Reflection and Privacy-Preserving Collection of Computer-Use Behaviors} 

% \title{Towards Longitudinal Alignment via Contextual Reflection and Privacy-Preserving Behavioral Data Collection}

\title[Towards Longitudinal Human-LLM Alignment]{Stayin' Aligned Over Time: Towards Longitudinal Human-LLM Alignment via Contextual Reflection and Privacy-Preserving Behavioral Data}
\author{Simret Araya Gebreegziabher}
\affiliation{
  \institution{University of Notre Dame}
  \city{Notre Dame}
  \state{IN}
  \country{United States}
}
\email{sgebreeg@nd.edu}

\author{Allison E Sproul}
\affiliation{
  \institution{University of Notre Dame}
  \city{Notre Dame}
  \state{IN}
  \country{United States}
}
\email{asproul@nd.edu}

\author{Yinuo Yang}
\affiliation{
  \institution{University of Notre Dame}
  \city{Notre Dame}
  \state{IN}
  \country{United States}
}
\email{yinooyang@nd.edu}

\author{Chaoran Chen}
\affiliation{
  \institution{University of Notre Dame}
  \city{Notre Dame}
  \state{IN}
  \country{United States}
}
\email{cchen25@nd.edu}

\author{Diego Gómez-Zará}
\affiliation{
  \institution{University of Notre Dame}
  \city{Notre Dame}
  \state{IN}
  \country{United States}
}
\email{dgomezara@nd.edu}

\author{Toby Jia-Jun Li}
\affiliation{
  \institution{University of Notre Dame}
  \city{Notre Dame}
  \state{IN}
  \country{United States}
}
\email{toby.j.li@nd.edu}


\begin{abstract}
Current human-AI alignment and evaluation methods for large language models (LLMs) often rely on preference signals collected immediately after an interaction. This practice implicitly treats preference as static, even though many LLM-mediated decisions unfold over time and may be re-evaluated differently after real-world consequences and observed outcomes. Therefore, we argue for a methodological shift from single-moment preference elicitation to longitudinal, context-situated alignment measurement. \revision{We present a methodological framework and system for longitudinal preference collection} that combines (1) in-situ evaluation, (2) context-triggered follow-up reflection, and (3) privacy-preserving behavioral traces that help contextualize evaluation revisions. As an instantiation of this methodology, we introduce \toolname{}, a browser-based system that detects consequential LLM interactions, prompts reflection across later decision points, and supports progressive, user-controlled consent for sharing behavioral data. Through a two-week proof-of-concept deployment with eight participants, our approach captured differences between participants’ immediate and delayed evaluations of LLM outputs across dimensions including accuracy, trust, and relevance. Our findings highlight the limitations of single-moment preference datasets and underscore the importance of longitudinal methods for alignment evaluation in everyday use.
\end{abstract}



\begin{teaserfigure}
    \centering
    \includegraphics[
    width=0.8\linewidth,
    trim={0 13cm 0cm 0},
    clip
    ]{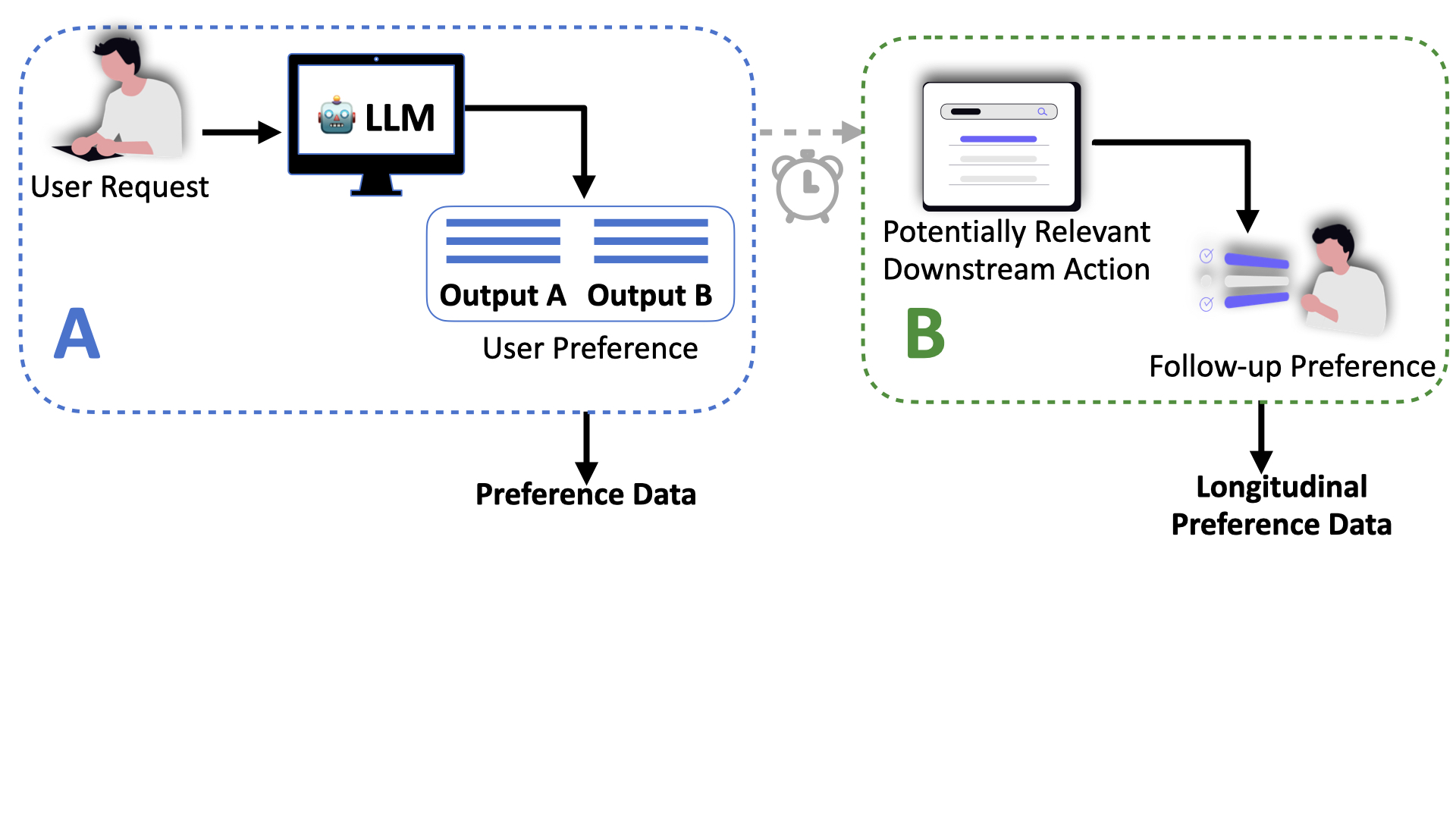}
    \caption{\revision{(A) Conventional approaches collect preference feedback during or immediately after an LLM interaction, producing a single-moment preference signal. (B) \toolname{} extends this measurement window by eliciting a follow-up reflection after detecting a potentially related downstream event. Together, the initial and follow-up evaluations provide temporally grounded longitudinal preference data. The downstream event serves as a proxy for triggering reflection and does not establish that an action occurred or was caused by the LLM interaction.}}
    \Description{A two-panel flow diagram contrasting conventional preference elicitation with BITE. Panel A, enclosed by a blue dashed border, shows a user request flowing to an LLM, two candidate outputs, a user preference, and preference data. A clock and downward arrow indicate that time passes before Panel B. Panel B, enclosed by a green dashed border, shows a potentially related downstream event leading to a follow-up reflection and longitudinal preference data. The diagram illustrates that conventional approaches collect feedback at one moment, whereas BITE combines an initial evaluation with a later, context-triggered evaluation.}
    \label{fig:teaser}
\end{teaserfigure}

\maketitle

\section{Introduction}

Preference learning and human-AI alignment for large language models (LLMs) are largely grounded in feedback collected immediately after an interaction, as in reinforcement learning from human feedback (RLHF)~\cite{ouyang2022training} and other preference-optimization approaches~\cite{christiano2017deep,ziegler2019fine}. Human feedback provides useful signals about users' reactions at the moment of generation, but they are often collected before users have acted on an output, verified claims, or encountered consequences. This creates an upstream measurement gap in preference elicitation. Alignment pipelines receive limited evidence about how users assess LLM outputs after using them in context~\cite{gao2023cirs}. We address this gap with a methodological framework and browser-based system, \toolname{}, for collecting temporally and contextually grounded feedback after users act on LLM outputs.

Many consequential LLM-mediated activities unfold beyond the moment of interaction. Tasks such as choosing products, planning travel, or drafting important communications involve follow-through actions whose outcomes may only become apparent later. In these settings, users may reassess an output after encountering additional information or experiencing its consequences. Prior work similarly suggests that preferences can be constructed in context and may be sensitive to when and under what conditions they are elicited~\cite{slovic1995construction,schwarz1999reports,loewenstein1996out}. Optimizing only for immediate feedback may therefore miss criteria that users apply after downstream use or overemphasize short-term satisfaction~\cite{gao2023cirs}.

This gap shows a methodological limitation in current alignment pipelines. Current user preference elicitation approaches capture proximal reactions but under-represent temporally grounded preference signals and changes. We argue that existing alignment methods are shallow in time. While effective for single-turn judgments, they provide limited visibility into how user preferences and evaluations may evolve over time. Because human preferences are time-inconsistent and sensitive to when evaluations are made~\cite{loewenstein1992anomalies}, immediate judgments may not reflect later assessments after outcomes unfold. \revision{If delayed judgments systematically diverge from immediate ones, optimizing on one-shot feedback may overestimate practical alignment. This temporal scope differs from multi-turn reinforcement learning, which generally extends the optimization horizon across conversational turns or modeled rollouts~\cite{shani2024multi}. Conversely, we focus on what happens after the LLM interaction takes place and users browse elsewhere and take actions according to the LLM outputs.}

Closing this gap is methodologically challenging. Longitudinal measurement in real-world settings requires capturing interactions and downstream behaviors over time, across technological environments and contexts, while minimizing user burden and preserving privacy. Prior work in experience sampling and HCI in-the-wild methods highlights the tension between ecological validity and intrusive data collection \cite{csikszentmihalyi1987validity,brown2011into}. Moreover, understanding how and why users reassess LLM outputs requires contextual signals about their subsequent actions and decisions, rather than repeated preference elicitation alone. However, collecting such behavioral traces introduces significant privacy concerns~\cite{fang2026ai}.

To address these challenges, we introduce \toolname{}, a browser-based system for capturing temporally grounded alignment signals in everyday LLM use. \toolname{} combines three key components: \textbf{(1) in-situ preference elicitation that captures immediate evaluations of LLM interactions, (2) outcome-aware follow-up preference elicitation triggered by downstream actions, and (3) privacy-preserving behavioral traces collected through a progressive consent model}. By linking LLM interactions to later user behavior and reflection, our approach enables the study of alignment as a temporal process.

We deploy \toolname{} in a two-week in-the-wild study with 8 participants and 182 captured LLM conversations \revision{71 of which had followup reflections}. In the study, we observe three longitudinal patterns after users act on model outputs. First, immediate and delayed judgments diverged most on trust and relevance, indicating that our participants re-evaluated LLM outputs after downstream use. Second, preference updates were directionally asymmetric. Trust changed at follow-up ($p = 0.045$), with most trust revisions moving downward. Participants attributed these revisions to outcome verification and context reconstruction (e.g., checking external sources, revisiting options), rather than passage of time alone. \revision{These exploratory findings demonstrate that \toolname{} can capture delayed reassessments that are not observable through single-moment preference elicitation and suggest that some evaluations may be temporally contingent.}

\revision{In summary, this paper makes three contributions:}
\begin{itemize}
    \item \revision{\textbf{A methodological framework} for collecting temporally and contextually grounded evaluations by connecting in-situ preference elicitation, context-triggered follow-up reflection, and downstream behavioral data.}

    \item \revision{\textbf{\toolname{}, a privacy-aware browser-based system} that operationalizes this framework through cross-context activity detection, outcome-aware follow-up prompts, and progressive, user-controlled consent for sharing browsing traces.}

    \item \revision{\textbf{Exploratory findings from a two-week in-the-wild deployment} demonstrating that \toolname{} can capture delayed reassessments after users act on LLM outputs and identifying dimensions and contextual processes associated with those reassessments.}
\end{itemize}
\section{Related Work}

\subsection{RLHF and Preference Learning}
A central premise in reinforcement learning from human feedback (RLHF) is that human preferences can be elicited as stable, immediately evaluable judgments over model behavior. The canonical formulation in \citet{christiano2017deep} operationalizes feedback as rapid, local pairwise comparisons that are largely decontextualized from downstream use. Contemporary RLHF pipelines also retain this structure. For example, \citet{ouyang2022training} relies on short textual judgments as the primary training signal, rather than measurements of how model outputs perform when acted upon in real-world settings.

While this design enables scalable data collection, it also constrains what is being optimized. Because feedback is collected at interaction time, RLHF systems are implicitly trained to optimize for short-term, surface-level signals that correlate with immediate approval~\cite{gao2023cirs}. Prior work has shown that such proxy signals can be exploited during optimization, leading to the amplification of stylistic features that do not necessarily reflect deeper alignment with user goals \cite{ziegler2019fine}. More broadly, the reward-hacking and specification-gaming literature highlights how optimizing imperfect proxies for human preferences can diverge from what users would endorse once outcomes unfold over longer horizons \cite{amodei2016concrete}. 

Our work builds on this line of critique by focusing on a specific and underexplored dimension of temporality. When the consequences of model outputs are delayed, or when evaluation depends on contextual information revealed only after follow-through, one-shot judgments collected at interaction time may provide an incomplete or even misleading signal of alignment~\cite{benk2026same}. This concern aligns with prior observations that humans struggle to accurately evaluate options in isolation without experiencing their downstream effects \cite{irving2018ai}. These limitations suggest that preference signals used in current alignment pipelines may not fully capture how user evaluations evolve over time, motivating methods that account for temporally situated judgments.

\subsection{Short-term vs. Long-term Human Judgment}
A substantial body of work in psychology and behavioral economics shows that human judgments are not stable across time, but instead depend on when and how they are elicited. Prior research on time inconsistency and preference construction demonstrates that evaluations formed in the moment can differ systematically from those formed after reflection, additional information, or lived consequences \cite{laibson1997golden,odonoGhuRabin1999doing,slovic1995construction}. Rather than revealing fixed preferences, momentary judgments often reflect context-dependent constructions that may be revised as individuals gain new information or experience outcomes.

One key mechanism underlying these differences is the distinction between immediate and reflective evaluation. Immediate judgments tend to rely on salient, readily accessible cues, such as fluency or coherence, particularly when decisions are made quickly or in isolation \cite{alter2012fluency,tversky1974judgment}. In contrast, reflective judgments can incorporate additional considerations and follow-through, including downstream outcomes, opportunity costs, and alignment with longer-term goals \cite{gilbert2002immune}. These later evaluations are often informed by contextual information unavailable at the time of the initial judgment. These findings have important implications for alignment and preference learning. If evaluations depend on temporal context, then preference signals collected at a single moment may not reflect users' considered judgments once they act on model outputs. In particular, feedback elicited immediately after interaction may overrepresent surface-level qualities while underrepresenting outcome-based considerations that only emerge over time. This suggests that alignment signals are inherently temporally situated, motivating approaches that capture how evaluations evolve alongside user experience.

% Previous work in psychology, behavioral economics, and HCI suggests that human judgments are often time-inconsistent \cite{laibson1997golden,odonoGhuRabin1999doing,slovic1995construction}. Preference evaluations made in the moment can differ from evaluations made after reflection, additional information, or lived consequences \cite{loewenstein1996out,gilbert2002immune,slovic1995construction}. Immediate judgments tend to privilege salient and readily accessible cues, such as fluency, and readability \cite{alter2012fluency,tversky1974judgment}, whereas delayed judgments can incorporate opportunity costs, downstream outcomes, and values-based reasoning that only becomes available once a recommendation has been acted upon or reconsidered \cite{loewenstein1996out,gilbert2002immune,odonoGhuRabin1999doing}.

\subsection{Web browsing and Behavioral Traces}

% \todo{Refactor subsection}
% - how can we do user behavior modeling using web browsing data?
% - how to connect an LLM chat recommendation to later behaviors like booking, purchasing, emailing etc

Prior research has long treated interaction history as a meaningful behavioral signal. Early work on ``read wear'' and ``edit wear'' showed that accumulated traces can reveal attention, task progress, and likely future actions in digital artifacts~\cite{hill1992editwear}. Beyond recording behavior, these traces act as memory cues that help users reflect on prior actions and changes over time. Similarly, prior work on social translucence argued that making behavioral traces visible to users can improve coordination and accountability in collaborative settings \cite{erickson1999socially,erickson2000social}. More recent work extends this idea to recommendation settings by leveraging personal digital traces (e.g., calendars, communication logs, and activity histories) to infer latent preferences and support group-level decisions \cite{wei2016grouplink}.

In our setting, these prior studies suggest a modeling strategy for treating web browsing and chat interaction logs as temporally structured evidence of evolving user intent. Instead of using a single immediate preference label, a model can aggregate multi-stage signals such as query reformulations, browsing patterns, dwell time, abandonment, and follow-through (e.g., moving from chat to booking, payment, or email confirmation). Studies of everyday browsing tab management further indicate that browsing behavior often encodes deferred commitments and pending tasks, not only instantaneous interest~\cite{chang2021tab}.

This perspective is central to evaluating recommendations generated by LLMs because a response that receives high in-chat ratings may still yield poor downstream utility, while a less impressive initial response may better support longer-horizon success. Large-scale in-the-wild chats reveal substantial behavioral diversity~\cite{zhao2024wildchat,chen2026digitalcoach}, while fine-grained interaction signals such as mouse trajectories provide complementary cues for modeling user intent \cite{zhang2024dismouse}. At the same time, real-world deployments underscore the importance of addressing profiling and privacy risks in trace-based systems \cite{vekaria2025big,fang2026ai}.

Work on end-user web automation likewise shows that intent is often realized through multi-step, cross-interface workflows, motivating delayed and outcome-aware evaluation rather than one-shot judgment \cite{leshed2008coscripter,chen2026taskartisan}. Complementary reviews and domain trials similarly argue for tying recommendations to downstream consequences instead of immediate reactions \cite{niu2026ethical,tao2026care}. Recent educational deployments also demonstrate that longitudinal traces capture users' evolving engagement with planning, reflection, and collaborative discussion affordances, not just response-level quality \cite{yang2026lessonsrealworlddeploymentcognitionpreserving,peng2025glitter,yang2025spark,zhang2026editrail,yang2024vizcode}. Building on these findings, we argue that alignment-oriented feedback pipelines should explicitly connect recommendation events to later behavioral outcomes through longitudinal, cross-context traces.

\section{Design Goals}
\label{sec:design_goals}
The design of \toolname{} was guided by four overarching goals including capturing temporally situated reflections on LLM interaction, preserving meaningful user control over personal data, and minimizing disruption to users' natural workflows. These goals draw on established principles from human–computer interaction and privacy research.

% what counts as meaningful consent

\subsection{\revision{\textbf{DG1: Capturing Temporally Separated Reassessments }}}

A key limitation in current evaluations of LLM systems is that user feedback is often collected immediately after interaction, before users have experienced the downstream consequences of LLM suggestions. However, research in reflective interaction design and experience-centered sampling suggests that users' interpretations of technology often evolve as they integrate system outputs into real-world contexts~\cite{sengers2005reflective}.

To address this gap, the system must capture both immediate and delayed reflections on LLM-generated content. This design goal draws inspiration from Schön's theories of reflection-in-action and reflection-on-action~\cite{munby1989reflection}, which distinguish between immediate responses during an activity and retrospective evaluations after the activity has unfolded. By linking LLM interactions to subsequent user behavior, the system can study how perceptions of LLM interaction may shift over time.

\subsection{\textbf{DG2: Enable Context-Aware Reflection Without Burdening Users}}

Continuous surveys or diary studies can impose a significant burden and disrupt users' workflows. Prior work in experience sampling and context-aware computing suggests that prompts are most effective when they are situated within meaningful events rather than delivered at arbitrary times~\cite{csikszentmihalyi1987validity}. 

Therefore, the system must prompt users to reflect on and express their preferences when interactions occur in decision domains that may lead to consequential outcomes. This event-driven design may minimize unnecessary interruptions while increasing the likelihood that reflections occur at moments when users can meaningfully assess the impact of the LLM response.

\subsection{\textbf{DG3: Support Progressive and Contextual Consent}}

Collecting users' LLM interaction and browsing data raises significant privacy concerns~\cite{zhang2024s, vekaria2025big}. To that end, \citet{nissenbaum2004privacy}'s Contextual Integrity (CI) theory argues that privacy is not a static state of secrecy but rather a set of norms in which data sharing is contingent on the specific context, purpose, and method of collection. Recent work has argued for using AI to go beyond requesting broad permissions upfront. In other words, the system adopts a progressive consent model in which users are asked to share contextual browsing data only when it becomes relevant~\cite{slate2025iterative}. To support this, the system should make the scope and purpose of data collection and sharing explicit, allowing users to make deliberate, real-time decisions about their data.

\subsection{\textbf{DG4: Study LLM Use in Natural Environments}}
Many studies of LLM decision support rely on short tasks that may not capture how users integrate LLM interactions and suggestions into everyday activities. Research in in-the-wild HCI methods suggests that studying technology in natural contexts can reveal behaviors and outcomes that would not likely emerge in controlled environments~\cite{brown2011into}. The system, therefore, should operate unobtrusively in the background during users' regular browsing activities and trigger prompts only when relevant events occur. This approach enables the study of LLM interaction and reflection over longer periods as they naturally occur within users' daily workflows.

\begin{figure*}
    \centering
    \includegraphics[
    width=0.75\linewidth,
    trim={0 2cm 0 0},
    clip
    ]{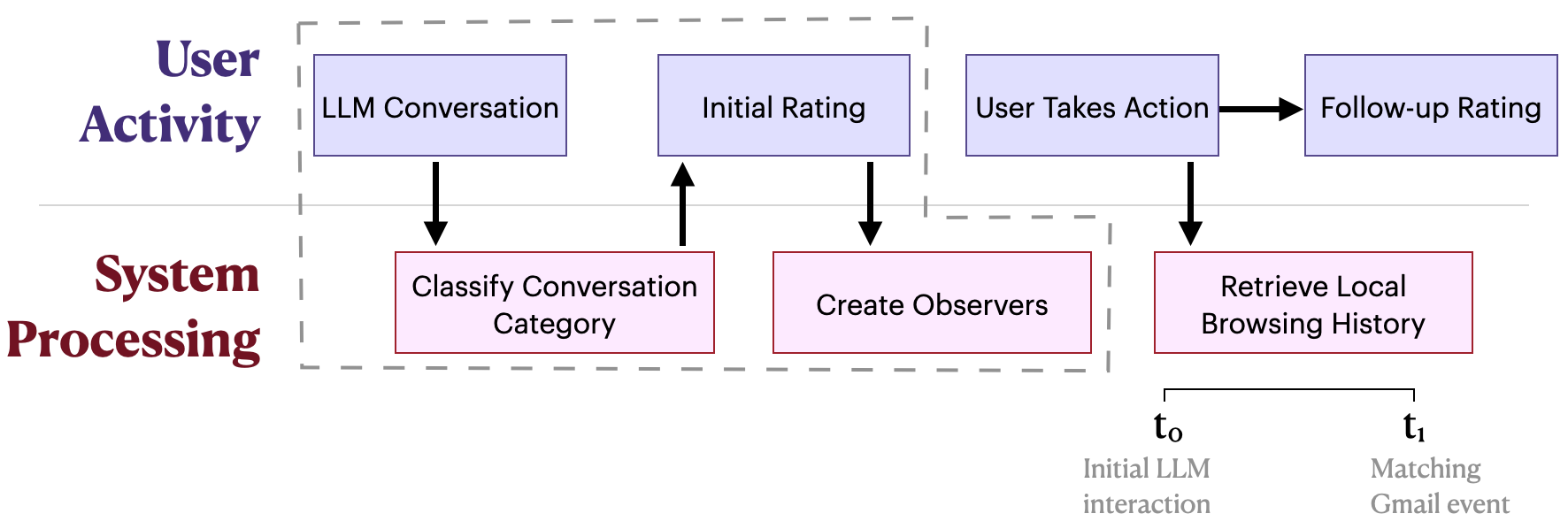}
    \caption{Workflow of the \toolname{} system. Following an LLM conversation, users provide an initial rating. The system classifies the interaction and instantiates observers to track downstream actions. Upon detecting a relevant real-world outcome (e.g., via browsing or email signals), the system retrieves contextual data and prompts a follow-up rating, supporting longitudinal analysis of alignment.}
    \Description{A two-row workflow diagram showing the longitudinal reflection pipeline of the BITE system. The top row, labeled User Activity, shows a sequence of stages: LLM conversation, initial rating, user takes action, and follow-up rating. The bottom row, labeled System Processing, shows corresponding system steps beneath each stage: classify conversation category, create observers, and retrieve local browsing history. Arrows connect user actions and system processes to illustrate how an initial LLM interaction is classified, monitored over time, and later linked to downstream actions and follow-up evaluation.}
    \label{fig:system-overview}
\end{figure*}

\begin{figure}[!htb]
    \centering
    \includegraphics[trim={16cm 0 20cm 0},clip,width=\columnwidth]{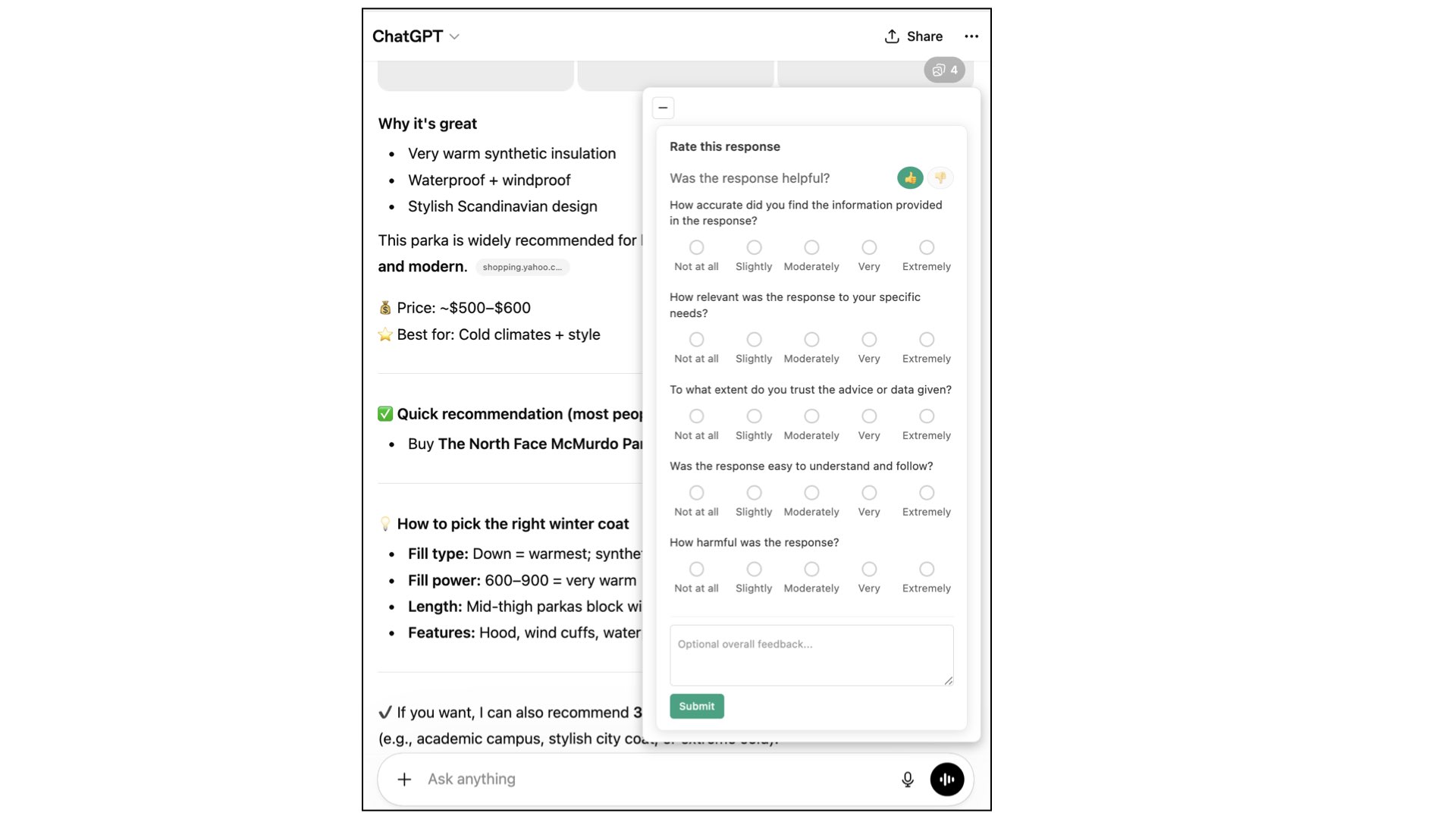}
    \caption{In-situ rating interface for immediate feedback. After an LLM response, users provide a one-shot evaluation of helpfulness, accuracy, relevance, trust, and other dimensions, capturing judgments at the time of interaction.}
    \Description{Screenshot of the immediate feedback pop-up interface overlaid on a ChatGPT conversation window. The pop-up presents a Likert-scale questionnaire asking users to rate the response on helpfulness, accuracy, relevance, trust, ease of understanding, and harmfulness, with options ranging from ``Not at all'' to   ``Extremely''. A text box for optional feedback and a submit button appear at the bottom.}
    \label{fig:chatgppt_rating_screenshot}
\end{figure}

\section{The \toolname{} System}

Based on the above design goals, we developed \toolname{}, a Chrome extension for data collection and reflection that detects potentially consequential LLM interactions and solicits staged user consent to study downstream outcomes. In its current design, \toolname{} does not personalize, fine-tune, or update the underlying LLM, nor does it determine which rating represents the ``correct'' preference. Instead, it captures users' evaluations at different moments and in different contexts through a multi-stage interaction pipeline.

\subsection{User Scenario}
Consider Alice, a graduate student planning a summer trip. While using ChatGPT, Alice asks for recommendations on affordable destinations and receives suggestions for several cities. While Alice is interacting with the LLM chatbot, \toolname{} detects that the conversation relates to the \textit{travel} category, which is one of the decision domains the system tracks. After the conversation ends, a small pop-up appears (Fig~\ref{fig:chatgppt_rating_screenshot}) asking Alice to briefly rate her interaction with the LLM. The pop-up includes questions on how useful, clear, harmless, relevant, accurate, and trustworthy the interaction was through a 5-point Likert scale and open-ended text response (details in Appendix~\ref{app:response_eval_questions}).

\begin{figure*}[tbh]
    \centering
    \includegraphics[width=0.9\linewidth]{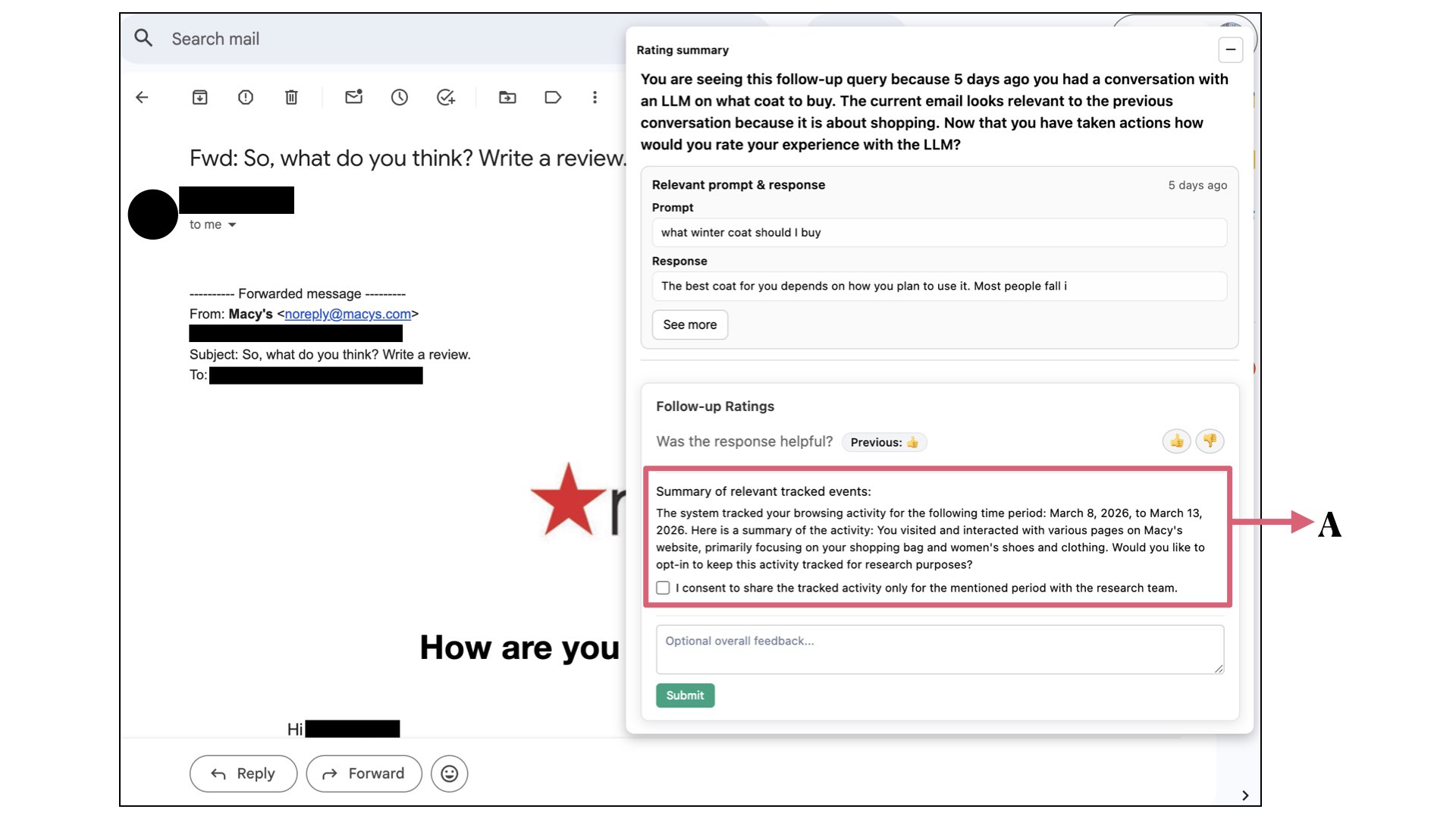}
    \caption{Follow-up rating interface triggered by a real-world event. When a relevant outcome is detected (e.g., an email related to a prior LLM-assisted decision), the system surfaces the original prompt and response, summarizes related user activity, and elicits a delayed evaluation.}
    \Description{Screenshot of the delayed follow-up rating interface displayed alongside a Gmail message window. The interface shows a summary of the earlier LLM interaction, including the original prompt and response, followed by a follow-up evaluation section. A highlighted consent box summarizes tracked browsing activity within a specific time window and asks whether the user consents to share that activity for research purposes. The figure emphasizes the event-triggered reflection and progressive consent mechanism.}
    \label{fig:email_rating_screenshot}
\end{figure*}

After interacting with the LLM, Alice browses different booking sites and books her travel itinerary. Alice then gets a booking confirmation through her email account. The extension detects that the email content falls within the same travel category and is similar to her previous LLM interaction. This triggers a follow-up reflection prompt (Fig~\ref{fig:email_rating_screenshot}). The follow-up prompt asks Alice whether the earlier LLM conversation influenced her decision and whether she would rate the previous interaction any differently after taking action.

At this point, the extension also asks Alice if she would like to share the browsing activity that occurred between the original LLM interaction and the email event (Fig~\ref{fig:email_rating_screenshot}-A). The interface clearly shows the time window and explains that the browsing data is stored locally in her browser unless she consents to share it. Alice can choose whether to share this information or keep it private. If she consents, only the browsing history within the specified time range is uploaded for research purposes. Through this staged interaction, \toolname{} captures Alice's initial perception of the LLM interaction, her subsequent real-world action, and her retrospective evaluation, while allowing her to retain control over which contextual data is shared.

\subsection{Key Features}

Guided by the design goals described in Section~\ref{sec:design_goals}, we implemented several key features in \toolname{} that enable temporally grounded reflection on LLM interaction while preserving user agency and minimizing disruption to everyday workflows (see Figure!\ref{fig:system-overview}.

\begin{table*}[t]
    \centering

    \begin{tabular}{p{0.18\textwidth}p{0.74\textwidth}}
        \toprule
        \textbf{Topic} & \textbf{Examples} \\
        \midrule
        \textit{shopping} & Product recommendations, what to buy, prices, deals, gifts, reviews, clothing/footwear/fashion \\
        \textit{job\_career} & Job applications, resume/cover letter help, career advice, grad school applications, interviews \\        \textit{travel} & Trip planning, destinations, flights, hotels, itineraries, travel recommendations \\
        \textit{homework} & Homework help, assignment support, tutoring, studying, exam prep \\
        \textit{email\_drafting} & Drafting emails, composing messages, professional correspondence \\
        \textit{relationship} & Personal relationship advice, dating, family, friendship, interpersonal issues \\
        \textit{restaurant} & Restaurant recommendations, dining suggestions, reservations, food spots \\
        \textit{fitness} & Exercise, workout plans, nutrition for fitness, health routines \\
        \textit{productivity} & Time management, organization, task management, productivity tips \\
        \bottomrule
    \end{tabular}
    \caption{Tracked topic categories used for event-triggered reflection prompts.}
    \Description{Table listing the tracked topic categories used to trigger event-based reflection prompts in the BITE system. The first column contains nine categories: shopping, job_career, travel, homework, email_drafting, relationship, restaurant, fitness, and productivity. The second column provides example interaction types for each category, such as product recommendations for shopping, trip planning for travel, and assignment support for homework.}
    \label{tab:tracked-topics}
\end{table*}

\subsubsection{\textbf{Event-Triggered Reflection Prompts}}

To support temporally grounded preference elicitation (DG1) while avoiding unnecessary interruptions (DG2), \toolname{} uses an event-driven mechanism that activates only when an LLM interaction is likely to lead to downstream action. \revision{This design is informed by Dewey's distinction between ordinary, immediate judgment and reflective thought. For Dewey, reflection becomes especially important when a situation involves uncertainty and when the consequences of an initial judgment are not yet fully known~\cite{dewey2022we}. \toolname{} does not treat every LLM interaction as equally consequential, rather it identifies interactions for which later action may provide additional grounds for reassessment.}

\revision{We determine that an LLM interaction is likely to lead to action using a set of predefined topics shortlisted by the researchers (Table~\ref{tab:tracked-topics}). When the user is interacting with the LLM, \toolname{} prompts a locally deployed Meta Llama model (Prompt in Appendix~\ref{app:topic_classification_prompt}) to categorize the conversation into one of the nine tracked topics. The tracked topics are not intended to be exhaustive, but instead provide a proof-of-concept set of domains in which an LLM response may shape a subsequent decision or action.} Additional topics could be incorporated as the system scales.

While the user is interacting with the LLM, \toolname{} prompts a locally deployed Meta Llama model to categorize the conversation into one of nine tracked topics established in the current implementation. If a conversation is classified into one of these topics, the system displays an initial appraisal prompt through a pop-up on the user's screen (Fig.~\ref{fig:chatgppt_rating_screenshot}). The pop-up asks users to rate the usefulness, trustworthiness, accuracy, and perceived relevance of the LLM response. Users may dismiss, minimize, or ignore the prompt without penalty, allowing participation to remain user-initiated rather than mandatory.

\revision{This initial prompt is not intended to constitute reflection in the full Deweyan sense. Instead, it captures the user's provisional evaluation at the moment of interaction, before the consequences of acting on the response are known. Anchoring this appraisal to the original interaction provides a contemporaneous record against which later judgments can be compared. The event-triggered design also reduces unnecessary prompting during low-consequence interactions while prioritizing situations in which subsequent experience may provide meaningful grounds for reconsideration.}

\subsubsection{\textbf{Outcome-Aware Follow-Up Reflection}}

To capture retrospective reflection after users have acted on LLM interaction (DG4), \toolname{} detects potential downstream actions that occur after the initial interaction. As a proof-of-concept implementation, the current system uses Gmail activity as a proxy for an event that may relate to a prior LLM interaction, rather than as evidence that a particular action occurred, produced a consequence, or was caused by the LLM interaction.

\toolname{} uses a two-step process to determine whether a follow-up reflection prompt should be shown. Email topics are collected through a Gmail content script from the rendered message view in the active browser tab. We currently support Gmail because these DOM selectors were implemented and validated against Gmail's interface; extending to other providers would require provider-specific DOM mapping and re-validation of extraction reliability. First, the system applies the same topic classification pipeline used during the initial interaction to determine whether the current email belongs to one of the tracked decision domains. Only emails classified into a tracked topic are considered candidate downstream events. Second, for candidate emails, the system compares the email content with previously recorded LLM interactions on the same topic using lexical overlap. Specifically, we compute the Jaccard similarity between the tokenized word sets of the email text and the earlier LLM interaction content. This produces a similarity score between 0 and 1, where higher values indicate greater semantic overlap. We selected Jaccard similarity as a lightweight on-device matching heuristic that supports local on-browser computation. If the similarity score exceeds 0.5, \toolname{} interprets the email as a likely follow-through action related to the earlier interaction and triggers a delayed follow-up reflection prompt.

\subsubsection{\textbf{Progressive Consent for Contextual Data Sharing}}
To support understanding how users move from an initial LLM interaction to a committed downstream action while preserving privacy (DG3), \toolname{} adopts a progressive consent mechanism for sharing contextual browsing data. Rather than requesting broad access to users' browsing history upfront, the system stores browsing activity locally on the device and defers data-sharing decisions until contextual information becomes relevant to a specific follow-up event. The locally retained trace includes visited domains, timestamps, and page titles.

During the follow-up reflection stage, \toolname{} presents the user with the bounded time window between the original LLM interaction and the detected downstream email event. At this point, users are asked whether they consent to sharing their browsing activity for this specific interval, only for relevant domains associated with the event. This event-bound consent model ensures that data sharing is limited in scope, temporally bounded, and directly tied to an explicitly communicated purpose. This aligns with principles of contextual integrity~\cite{nissenbaum2004privacy} by allowing users to make situated privacy decisions when the relevance of the requested data is clear. Users retain full control to approve or decline sharing on a per-event basis.

\subsection{Implementation}

\toolname{} is implemented as a Chrome browser extension using Manifest V3. The system consists of a background service worker, content scripts injected into supported web applications (e.g., LLM interfaces and Gmail), and lightweight pop-up interfaces for user reflection prompts. Content scripts monitor DOM events such as prompt submission, response rendering, and email composition, and serialize relevant interaction data into structured event objects. \toolname{} was implemented to work with three LLM-chatbot interfaces including OpenAI's ChatGPT\footnote{https://chatgpt.com/}, Claude\footnote{https://claude.ai/}, and Google Gemini\footnote{https://gemini.google.com/}, DOM structure.
The extension uses predefined selectors for message-thread containers and extracts only text from those targeted nodes in the currently loaded view.

These events are passed to the background service worker through Chrome's runtime messaging APIs, where the system coordinates domain classification, event matching, and local browsing trace management. For on-device inference, we use a locally deployed Meta Llama3.1 70B model to perform topic classification and generate summaries of contextually relevant browsing events in the backend.

\begin{table*}[tbh]
\centering
\resizebox{\textwidth}{!}{%
\begin{tabular}{@{}lllllllll@{}}
\toprule
\textbf{PID} & \textbf{Gender} & \textbf{Age} & \makecell[l]{\textbf{Education}\\ \textbf{Level}} & \textbf{Occupation} &
\makecell[l]{\textbf{How often do you} \\ \textbf{use LLMs?}} &
\makecell[l]{\textbf{How long have you} \\ \textbf{been using LLMs?}} &
\makecell[l]{\textbf{How often do you} \\\textbf{rely on LLMs when} \\ \textbf{making decisions?}} &
\makecell[l]{\textbf{How much do you trust} \\ \textbf{advice generated by LLMs?}} \\ \midrule
P1 & Female & 25-34 & Bachelor's  & PhD Student & Multiple times per day & More than 2 years & Rarely & Neutral \\
P2 & Male & 18-24 & Bachelor's  & PhD Student & About once per day & More than 2 years & Sometimes & Mostly trust \\
P3 & Female & 25-34 & Bachelor's  & Realtor & Multiple times per day & 1–2 years & Often & Mostly trust \\
P4 & Male & 18-24 & Master's  & Teaching Assistant & Several times per week & 1–2 years & Often & Mostly trust \\
P5 & Female & 18-24 & Some College & Student & About once per day & 1–2 years & Rarely & Neutral \\
P6 & Female & 18-24 & Bachelor's  & Student & About once per day & 1–2 years & Sometimes & Mostly distrust \\
P7 & Female & 25-34 & Bachelor's  & IT consultant & Multiple times per day & More than 2 years & Very Often & Completely trust \\
P8 & Female & 18-24 & Bachelor's  & Student & Several times per day & 1-2 years & Sometimes & Mostly trust \\\bottomrule
\end{tabular}
}
\caption{Study Participant Demographics}
\Description{Table summarizing the demographics and LLM usage patterns of eight study participants. Participants vary in age from 18 to 34 and include students, a PhD student, a teaching assistant, a realtor, and an IT consultant. Most participants report using LLMs daily or multiple times per day and have between one and more than two years of experience. Trust levels range from mostly distrust to complete trust, with most participants reporting moderate to high trust.}
\label{tab:participants}
\end{table*}

\section{Evaluation}

We conducted a deployment study\footnote{The study protocol has been reviewed and approved by the IRB at our institution.} to evaluate whether \toolname{} can effectively capture temporally separated reflections on LLM interaction while preserving user understanding and control over data sharing. We ask the following research questions:

\begin{itemize}
\item  \textbf{RQ1}: How do participants' evaluations of LLM responses change between immediate interaction-time judgments and later reflections?
\item  \textbf{RQ2}: What factors drive changes in participants' evaluations over time?

\end{itemize}

\subsection{Study Design}
We conducted a two-week-long longitudinal deployment study in which participants installed the \toolname{} Chrome extension and used it during their everyday interactions with LLM systems. The extension ran in the background and triggered prompts when relevant events occurred, including LLM conversations in tracked domains and subsequent email activity.

During the study, participants encountered an immediate reflection prompt following a detected LLM interaction and a follow-up reflection prompt triggered when the system detected a downstream email event within the same domain category. During the follow-up stage, participants were also asked whether they consented to share browsing activity within the bounded time window between the two events.

\subsection{Participants}
We recruited participants through social media postings. We targeted participants who regularly use LLM tools for their everyday tasks. During prescreening, we confirmed eligibility criteria required by the deployment setup. Participants had to be at least 18 years old, use a personal computer with Chrome as their primary browser, and actively use Gmail, since downstream follow-up detection is triggered from Gmail activity.

The final sample included eight participants with varied backgrounds (Table~\ref{tab:participants}). Participants installed the browser extension on their personal computers and used it during their normal workflows for the duration of the study.

\subsection{Procedure}
Each participant completed a structured onboarding session on Zoom before the deployment began. During onboarding, the research coordinator introduced the study goals, confirmed that participants were working on a personal computer with a Chrome browser, and administered a prescreening survey. Following that, the participants were provided a live demo of the extension (or an equivalent slide-based walkthrough), reviewed the informed consent document, and installed the \toolname{} extension.

After onboarding, participants used \toolname{} for two weeks in their normal routines. The extension logged eligible LLM interactions and prompted in-situ ratings, followed by downstream reflection prompts when related events were detected. In parallel, participants completed a daily diary survey that asked them to review the extension's recent-activity summary, report what they used an LLM for that day, and report their perceived trust, reliance, and satisfaction with \toolname{}. Participants were compensated with a base payment of 40 USD, with performance-based bonuses for consistent check-ins, and substantive free-response completion up to a total cap of 100 USD.

\subsection{Analysis}

To evaluate the effectiveness of the approach and \toolname{}, we analyzed both system-level metrics and participants' subjective responses. For quantitative analysis, we measured topic coverage across domains and compared immediate versus follow-up ratings to assess changes in participants' evaluations over time. To characterize directional update patterns, we also computed a Directional Asymmetry Index (DAI) for each dimension, defined as $DAI=(N_{up}-N_{down})/(N_{up}+N_{down})$, where $N_{up}$ and $N_{down}$ are the counts of upward and downward rating revisions between immediate and follow-up judgments. A positive DAI indicates more upward than downward revisions, a negative DAI indicates the reverse, and values near zero indicate no clear directional tendency. For qualitative analysis, we conducted a thematic analysis of interview data. Two researchers first independently coded two interview transcripts, compared code assignments iteratively, and resolved disagreements through discussion to establish a shared codebook. One researcher then applied the finalized codebook to the remaining transcripts. The resulting codes were synthesized into themes about participants' experiences with reflection, decision revision, and privacy and consent preferences while using \toolname{}.

\subsection{Findings}

Across the two-week deployment, \toolname{} \revision{captured 182 LLM conversations from eight participants. Out of the 182 conversations, 71 had both initial and follow-up ratings. Follow-up evaluations occurred an average of 44.0 hours after the initial evaluation ($SD=11.9$; $median=47.9$ hours).}
The collected interactions were concentrated in a subset of everyday-use domains, with homework and assignment-related use comprising most captured events (69\%), followed by shopping (9.6\%), productivity (7\%), and travel (5.8\%). Other categories, such as relationship, job\_career, and fitness, were less frequent. 

\revision{We first examine where immediate and follow-up evaluations diverged quantitatively. We then use participants' accounts and behavioral traces to explain why those revisions occurred. Our qualitative analysis identified outcome-driven revision, verification-driven revision, and context-reconstruction revision as reasons for our participants change in their ratings.}

\begin{table*}[bht]
\begin{tabular}{@{}lllllll@{}}
\toprule
Dimension & \# of changed ratings & Increased ratings & Decreased ratings & Mean $\Delta$ & DAI & Wilcoxon p \\ \midrule
Relevance & 54 & 18 & 36 & -0.32 & -0.33 & >.1 \\
Trust & 40 & 8 & 32 & -0.38 & -0.60 & .045 \\
Accuracy & 23 & 3 & 20 & -0.28 & -0.74 & >.1 \\
Clarity & 29 & 0 & 29 & -0.75 & -1.00 & .068 \\
Harmfulness & 15 & 15 & 0 & 0.21 & 1.00 & .083 \\ \bottomrule
\end{tabular}
\caption{Frequency, direction, and magnitude of changes between participants' initial and follow-up evaluations. $(\Delta = followUpRating - initialRating$) across evaluation dimensions. negative values indicate lower follow-up ratings and positive values indicate higher follow-up ratings.}
\Description{Table comparing changes between participants’ initial and follow-up ratings across five evaluation dimensions. Relevance changed in 54 cases, with 18 increases and 36 decreases; the mean change was -0.32 and the directional asymmetry index, or DAI, was -0.33. Trust changed in 40 cases, with 8 increases and 32 decreases; the mean change was -0.38 and the DAI was -0.60. Accuracy changed in 23 cases, with 3 increases and 20 decreases; the mean change was -0.28 and the DAI was -0.74. Clarity changed in 29 cases, all of which were decreases; the mean change was -0.75 and the DAI was -1.00. Trust was the only dimension with a Wilcoxon p-value below .05, at .045. Overall, follow-up ratings generally decreased for relevance, trust, accuracy, and clarity, whereas perceived harmfulness increased.}
\label{tab:results_ratings}
\end{table*}

\subsubsection{\revision{Where did reassessment occur?}}

\revision{For each paired evaluation, we calculated the difference between the follow-up and immediate rating: }

\begin{equation}
\Delta = followUpRating - initialRating
\end{equation}

\revision{Positive delta values indicate that a rating increased at follow-up, while negative values indicate that it decreased. For harmfulness, a positive value indicates that the response was perceived as more harmful at follow-up. As shown in Table~\ref{tab:results_ratings},  the frequency and direction of reassessment varied across dimensions.} Relevance was reconsidered most frequently, with revisions occurring in both directions, whereas changes in trust, accuracy, and clarity were predominantly downward. 

Trust exhibited the clearest evidence of a systematic change. Among evaluations in which trust changed, most revisions reflected lower trust at follow-up ($DAI=-0.60$), and trust was the only dimension for which the difference between immediate and follow-up ratings reached statistical significance ($p=.045$). Relevance also changed frequently and was, on average, evaluated less favorably at follow-up, although its revisions were more heterogeneous and the aggregate difference was not statistically significant. Compared to the other dimensions harmfulness was revised less frequently and remained comparatively stable overall. However, of the limited observed harmfulness revision indicated greater perceived harm at follow-up. In this study, we observe that delayed reassessment was concentrated in particular interactions and dimensions, with trust providing the strongest quantitative evidence of a directional shift. \revision{Given the small, exploratory nature of our data, we interpret these quantitative patterns cautiously and draw on the qualitative findings below to contextualize what may have contributed to participants' revisions.}

\subsubsection{\revision{Why did participants revise their ratings?}}

In follow-up interviews, we found that preference shifts were driven by both immediate, outcome-based feedback and delayed, context-driven reminders. Immediate triggers occurred when participants executed LLM-generated code, tested ideas brainstormed with an LLM, or otherwise observed direct task outcomes shortly after interacting with the LLM. These triggers were often incidental, where participants revisited earlier advice after related conversations, Slack messages, or external reading prompted reconsideration of prior responses.

\paragraph{Outcome-driven revision}
\revision{Outcome-driven revisions occurred when participants changed their evaluation after attempting an LLM recommendation or observing its consequences. Participants described how the practical value of an LLM response sometimes became apparent only after they attempted to use it. In these cases, downstream experience provided evaluative evidence that was unavailable when the response was initially generated.}

These cases included executing LLM-generated code (P4), testing brainstormed ideas (P7), and using suggested recipes (P1). P1 illustrated this through an LLM-generated recipe. Although the response initially appeared useful, following it exposed limitations in its clarity and practical utility. P1 \textit{``did need a little help from YouTube''} because parts of the response were confusing and ultimately found that \textit{``the meal was not as much as I expected.''} While this specific interaction was not among the conversations rated through \toolname{}, it illustrates how an initially plausible response could receive a less favorable assessment after use. 

\revision{Outcome-based evaluation did not necessarily lead participants to reject the entire response. P2 partially acted on an LLM-generated travel itinerary, using its recommendations to identify attractions and hotels while modifying the proposed schedule and consulting other services. P2 explained, \textit{``I did act on it, mostly, like, not the entirety of it,''} describing the itinerary as useful for narrowing options but less dependable for time-sensitive information such as hotel prices.}

\paragraph{Verification-driven revision}
\revision{Participants also described reconsidering LLM responses after consulting other models, websites, applications, or people. Verification could challenge an initial evaluation, reinforce it, or leave the participant uncertain.} 

P3 provided the clearest example of a downward reassessment. P3 had relied on ChatGPT's description of a workplace policy and communicated it confidently over email to colleagues: \textit{``I was saying it, like, 100\%, because I was sure of the source.''} A colleague subsequently explained that the policy had been repealed, which P3 confirmed through another source and identified as a case of revised rating. P1 described a more routine form of corroboration when using LLM recommendations for travel and restaurants. Rather than relying on the generated response alone, P1 moved between ChatGPT, Google Maps, and a hotel-points application to compare ratings, locations, prices, and available options. In this case, external sources supplemented the LLM response and helped determine which recommendations were actionable.

\paragraph{Context reconstruction}

\revision{Participants' accounts also suggest that reassessment depended on their ability to reconstruct the original interaction. Specifically, what they had asked, which parts of the answer they used, and what happened afterward. Prompts, evaluation criteria, and retained conversation history provided different forms of support for this reconstruction. P1 described the study questions as prompting a broader reconsideration of LLM feedback. In particular, the repeated surveys made P1 more conscious of the LLM's tendency to respond in an overly affirming manner. This reflection led P1 to adopt a more critical stance toward apparently supportive responses and conclude that they \textit{``shouldn't take every feedback too high.''}} 

\begin{figure}
    \centering
    \includegraphics[width=\linewidth]{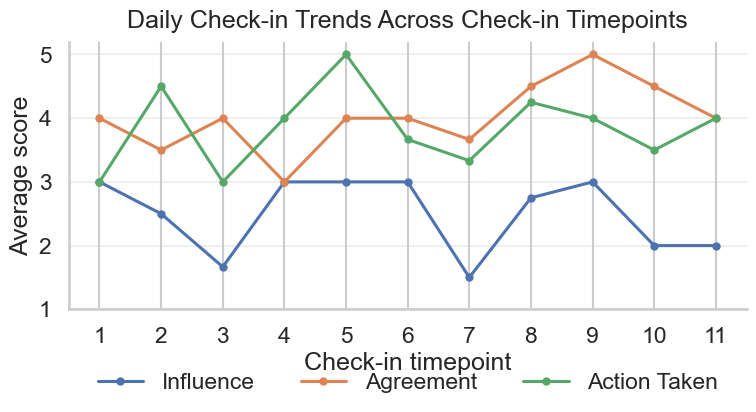}
    \caption{Average daily check-in scores across observed check-in timepoints. Influence reflects how much participants felt the LLM affected their thinking, agreement reflects how much they agreed with the LLM response, and action reflects whether they acted on the recommendation.}
    \Description{Line chart showing average daily check-in scores across eleven timepoints for three measures: influence, agreement, and action taken. The x-axis represents check-in timepoints and the y-axis shows average score from 1 to 5. Agreement and action taken remain consistently high over time, while influence fluctuates more substantially across timepoints.}
    \label{fig:daily_checkins}
\end{figure}

During the daily check-ins and follow-up interviews, participants reported mixed effects of the system on their own reflection. Some participants noted increased awareness (e.g., \textit{``It made me more aware of my AI use''}~(P8)), while others reported minimal impact (\textit{``Nothing too thought provoking today''}~(P2), suggesting that reflection is not uniformly triggered across interactions. Participants also noted challenges in reflecting during iterative interactions. For example, one participant stated, \textit{``If I do my discussion loop, it's hard to reflect,''~(P3)}. Most participants reported minimal concerns regarding data collection. However, a small number of responses indicated context-dependent sensitivity (e.g., \textit{``this research was a bit personal''~(P7)}), highlighting some potential variability based on the task at hand. It is also worth noting that, in some cases, participants reported that their LLM interactions occurred through mobile applications, suggesting potential gaps in coverage and challenges in capturing all relevant usage contexts. 

To complement our immediate and follow-up rating analyses, we examined participants' daily check-in responses over the two-week deployment (Figure~\ref{fig:daily_checkins}). Across the check-ins, agreement with LLM responses and reported action-taking remained consistently high, while perceived influence on participants' thinking exhibited greater temporal variability. Notably, agreement and downstream action remained elevated even at time points when perceived influence declined. For example, while influence ratings dropped at several later check-ins, participants continued to report relatively high agreement with responses and continued acting on recommendations.

\subsubsection{Browsing Traces as Supplemental Context}

Beyond changes in ratings, the collected browsing traces helped contextualize how participants moved from an initial LLM interaction to downstream action. In cases where participants consented to share bounded browsing activity, the traces surfaced possible comparison or verification activities. 

The collected browsing activity showed iterative comparison and verification behaviors, such as revisiting multiple pages, consulting external sources, or moving across booking websites before acting on an LLM recommendation. These traces helped contextualize why follow-up ratings sometimes diverged from participants' initial judgments. For instance, after discussing a school project with the LLM, P4 subsequently consulted several papers on arXiv and materials in Google Docs before revisiting their evaluation. In the follow-up rating, P4 noted, \textit{``I think with more time to sit with it, I have been able to reflect on its accuracy.''} P4 then revised the relevance and accuracy score from \textit{very relevant} to \textit{extremely relevant}.

Participants also described this contextual information as useful for reconstructing their own reasoning process. The browsing traces served as memory cues, helping them reflect on the information they consulted, the alternatives they considered, and where their understanding shifted over time. This suggests that behavioral context can complement delayed preference elicitation by functioning as contextual reminders during follow-up reflection. At the same time, participants highlighted the importance of granular control over what contextual data is shared. While most participants reported feeling comfortable with the consent mechanism, three participants expressed a desire for finer-grained sharing controls, such as selectively approving specific URLs. As one participant noted, they would \textit{``want to share some of these URLs''}~(P1), while another suggested that consent \textit{``can be domain specific''}~(P7).
\section{Discussion}
In this paper, we examined a method for eliciting user preference on LLM interaction over time, rather than only at the moment of output consumption. Across our deployment, we found that participants' immediate assessments and later reflections did not often align. This suggests that one-time ratings can miss meaningful changes in judgment that emerge after downstream action. We discuss implications for long-horizon alignment and evaluation, multi-source data collection, and consent-centered design.

\subsection{Long Horizon Alignment}
Our findings suggest that human-AI alignment should be evaluated as a temporal process. Immediate ratings remain useful for capturing first impressions, but they provide only a partial view of whether model guidance remains appropriate once participants act on that guidance in everyday settings~\cite{sengers2005reflective}. In our data, later reflections sometimes revised earlier judgments, indicating that temporally separated feedback can reveal misalignments that would be difficult to detect in single-session evaluations. Methodologically, this motivates complementing interaction-time measures with structured follow-up prompts tied to downstream events~\cite{csikszentmihalyi1987validity,brown2011into}. This does not replace existing evaluation practices; rather, it extends them by adding a longitudinal lens that better aligns with real-world decision trajectories.

The study also suggests that current LLM evaluation approaches provide an incomplete view of user preferences. Across our deployment study, participants frequently revised their initial ratings after acting on LLM responses, especially across dimensions such as accuracy and relevance. This indicates that immediate feedback may systematically overestimate model performance by rewarding surface-level qualities, such as fluency and coherence, that are more salient at the moment of generation but do not fully reflect downstream utility. 

Lastly, our results motivate the need for evaluation frameworks that incorporate temporally grounded signals. Rather than treating preferences as a static label, LLM evaluation frameworks should account for how preferences and judgments evolve as users integrate model outputs into real-world contexts. Concretely, this suggests augmenting existing benchmarks with delayed or outcome-aware measures, such as follow-up ratings tied to downstream actions, or trajectory-based evaluations that capture how users revise their decisions over time.

\subsection{Designing for Reflection-in-Action in Human-AI Alignment}
Our design implications align with prior work in reflective interaction design, where systems are built not only to support task completion but also to support users' reinterpretation of their own activity over time~\cite{sengers2005reflective}. In this framing, immediate prompts function as reflection-in-action checkpoints, while delayed prompts support reflection-on-action after consequences become visible~\cite{munby1989reflection}.

This interpretation also aligns with sensemaking of interaction, where users reinterpret earlier actions as new evidence emerges, rather than judging outputs as isolated artifacts. For alignment-oriented interface design, the implication is to treat preference elicitation as staged and situated to capture initial judgments in context, then provide lightweight re-entry points that help users connect outputs to downstream events. 

This design positions alignment as an ongoing human-AI process rather than a one-time annotation event. More explicitly, this reflects a bi-directional view of alignment, where adaptation occurs in both directions~\cite{shen2025bidirectional}. The LLMs may update based on user feedback and evolving judgments, while users themselves refine, reinterpret, and sometimes reconstruct their own preferences as they experience downstream consequences of AI-supported decisions. Rather than assuming that users hold stable preferences that can simply be extracted and optimized, our findings suggest that alignment emerges through mutual adaptation between human and system over time.

\subsection{Privacy-Aware Alignment Data Collection}

Collecting temporally grounded alignment signals requires access to behavioral context, such as the actions users take after interacting with an LLM. However, such data is inherently sensitive, raising important privacy concerns. Our findings suggest that users' willingness to share contextual data is not fixed, but contingent on how, when, and why the data is requested.

The progressive consent model implemented in \toolname{} allowed users to make context-specific decisions about data sharing when the relevance of that data became clear. Rather than granting broad permissions upfront, participants were asked to share bounded browsing activity only in relation to a specific LLM interaction and its downstream outcome. This approach aligns with theories of contextual integrity~\cite{nissenbaum2004privacy}, in which privacy expectations depend on the appropriateness of information flow within a given context.

Participants' responses indicate that such contextualized consent mechanisms can support a balance between data utility and user control. Users were more willing to share data when they understood its purpose and scope, and when it was clearly tied to a meaningful event. These findings suggest that future alignment data collection systems should move beyond static, one-time consent toward more dynamic, event-driven models that preserve user agency while enabling richer, longitudinal analysis.

\section{Limitations and Future Work}
Our study is subject to several limitations that point to important directions for future work. First, our deployment spans a two-week period, which limits the types of downstream outcomes we can observe. While this duration was sufficient to capture short-term preference revisions and immediate follow-through behaviors, many consequential decisions unfold over longer time horizons (e.g., career decisions, financial commitments, or long-term planning). As a result, our findings likely under-represent slower, more cumulative forms of preference change. Future work should explore longer-term deployments to examine how alignment judgments evolve over weeks or months, and whether early impressions continue to diverge from later evaluations over extended periods.

Second, our system relies on Gmail-based event detection as a proxy for downstream action. \revision{While we use this as a practical mechanism to link LLM interactions to real-world outcomes, it captures only a subset of user behaviors. Many downstream actions that may be taken in person, take longer than the two-week span, or abandoned may not be observable through our current approach. Consequently, we acknowledge that our measurement of ``actions taken'' is incomplete.} Future work could expand beyond email-based signals to incorporate additional sources of contextual evidence, such as cross-application activity, user-initiated annotations, or integration with other platforms, while continuing to prioritize user control and privacy.

Our participant sample is relatively small and skewed toward frequent LLM users. Participants who are already familiar with LLM systems may exhibit different reflection patterns, trust dynamics, and usage behaviors compared to less experienced or more skeptical users. Future studies should include more diverse populations, including non-expert users and domain-specific professionals, to better understand how longitudinal alignment manifests across different user groups and contexts. Additionally, while our study captures both immediate and delayed self-reported preferences, it does not directly measure objective task outcomes or long-term success. As such, we cannot fully disentangle whether preference revisions reflect improved judgment accuracy, changing expectations, or hindsight bias. Future work could combine subjective reflection with objective outcome measures to better understand how user perceptions align with actual performance over time.

Finally, our event-matching pipeline relies on topic classification and lexical overlap, which may miss semantically related events with low word overlap or incorrectly match events that share vocabulary but differ in intent. Future work could explore more robust local embedding-based matching while preserving privacy.

While \toolname{} is designed as a data-collection tool, it may also influence user behavior and reflection practices, creating social desirability bias \cite{grimm2010social}. The presence of prompts and the awareness of being studied may encourage participants to engage in more deliberate or critical evaluation than they otherwise would. This raises broader questions about how measurement interventions shape the phenomena they aim to observe. Future work could investigate how different prompting strategies affect user behavior, and explore more lightweight or passive approaches to capturing longitudinal alignment signals.

\section{Conclusion}
In this work, we argue that human-LLM alignment should be understood not as a static judgment but as a temporal process unfolding through interaction, action, and reflection. We introduce \toolname{}, a browser extension that captures both immediate and delayed user evaluations by combining in-situ prompts, outcome-aware follow-up reflection, and privacy-preserving behavioral context. In a two-week proof-of-concept deployment study, we found that participants' initial assessments of LLM outputs often diverged from their later reflections after taking downstream action. Our findings highlight potential limitations of single-instance preference signals and suggest the need for longitudinal, context-sensitive approaches to alignment evaluation. 

\begin{acks}
This work was supported in part by the Alfred P. Sloan Foundation (G-2024-22427), and an IBM Ph.D. Fellowship. Any opinions, findings, or recommendations expressed here are those of the authors and do not necessarily reflect the views of the sponsors.
\end{acks}

\balance

\bibliographystyle{ACM-Reference-Format}
\bibliography{sample-base}

@article{christiano2017deep,
  title={Deep reinforcement learning from human preferences},
  author={Christiano, Paul F and Leike, Jan and Brown, Tom and Martic, Miljan and Legg, Shane and Amodei, Dario},
  journal={Advances in neural information processing systems},
  volume={30},
  year={2017}
}

@article{ouyang2022training,
  title={Training language models to follow instructions with human feedback},
  author={Ouyang, Long and Wu, Jeffrey and Jiang, Xu and Almeida, Diogo and Wainwright, Carroll and Mishkin, Pamela and Zhang, Chong and Agarwal, Sandhini and Slama, Katarina and Ray, Alex and others},
  journal={Advances in neural information processing systems},
  volume={35},
  pages={27730--27744},
  year={2022}
}

@article{ziegler2019fine,
  title={Fine-tuning language models from human preferences},
  author={Ziegler, Daniel M and Stiennon, Nisan and Wu, Jeffrey and Brown, Tom B and Radford, Alec and Amodei, Dario and Christiano, Paul and Irving, Geoffrey},
  journal={arXiv preprint arXiv:1909.08593},
  year={2019}
}

@article{amodei2016concrete,
  title={Concrete problems in AI safety},
  author={Amodei, Dario and Olah, Chris and Steinhardt, Jacob and Christiano, Paul and Schulman, John and Man{\'e}, Dan},
  journal={arXiv preprint arXiv:1606.06565},
  year={2016}
}

@article{irving2018ai,
  title={AI safety via debate},
  author={Irving, Geoffrey and Christiano, Paul and Amodei, Dario},
  journal={arXiv preprint arXiv:1805.00899},
  year={2018}
}

@inproceedings{slate2025iterative,
  title={Iterative Contextual Consent: AI-enabled Data Privacy Contracts},
  author={Slate, Daniel D and Chen, Chaoran and Yao, Yaxing and Li, Toby Jia-Jun},
  booktitle={Proceedings of the 2025 Workshop on Human-Centered AI Privacy and Security},
  pages={84--91},
  year={2025}
}

@article{laibson1997golden,
  title={Golden Eggs and Hyperbolic Discounting},
  author={Laibson, David},
  journal={The Quarterly Journal of Economics},
  volume={112},
  number={2},
  pages={443--477},
  year={1997},
  publisher={Oxford University Press}
}

@article{odonoGhuRabin1999doing,
  title={Doing It Now or Later},
  author={O'Donoghue, Ted and Rabin, Matthew},
  journal={American Economic Review},
  volume={89},
  number={1},
  pages={103--124},
  year={1999}
}

@article{loewenstein1996out,
  title={Out of Control: Visceral Influences on Behavior},
  author={Loewenstein, George},
  journal={Organizational Behavior and Human Decision Processes},
  volume={65},
  number={3},
  pages={272--292},
  year={1996}
}

@article{slovic1995construction,
  title={The Construction of Preference},
  author={Slovic, Paul},
  journal={American Psychologist},
  volume={50},
  number={5},
  pages={364--371},
  year={1995}
}

@article{gilbert2002immune,
  title={The Trouble with Vronsky: Impact Bias in the Forecasting of Future Affective States},
  author={Gilbert, Daniel T and Pinel, Elizabeth C and Wilson, Timothy D and Blumberg, Stephen J and Wheatley, Thalia},
  journal={Journal of Personality and Social Psychology},
  volume={82},
  number={3},
  pages={353--366},
  year={2002}
}

@book{tversky1974judgment,
  title={Judgment under Uncertainty: Heuristics and Biases},
  author={Tversky, Amos and Kahneman, Daniel},
  year={1974},
  publisher={Science}
}

@book{alter2012fluency,
  title={Fluent Thinking: Why We Like What We Like and How We Think},
  author={Alter, Adam L},
  year={2012},
  publisher={Farrar, Straus and Giroux}
}

@chapter{schwarz1999reports,
  title={Reports of Subjective Well-Being: Judgmental Processes and Their Methodological Implications},
  author={Schwarz, Norbert and Strack, Fritz},
  booktitle={Well-Being: The Foundations of Hedonic Psychology},
  editor={Kahneman, Daniel and Diener, Ed and Schwarz, Norbert},
  pages={61--84},
  year={1999},
  publisher={Russell Sage Foundation}
}

@inproceedings{brown2011into,
  title={Into the wild: challenges and opportunities for field trial methods},
  author={Brown, Barry and Reeves, Stuart and Sherwood, Scott},
  booktitle={Proceedings of the SIGCHI conference on human factors in computing systems},
  pages={1657--1666},
  year={2011}
}

@article{csikszentmihalyi1987validity,
  title={Validity and reliability of the experience-sampling method},
  author={Csikszentmihalyi, Mihaly and Larson, Reed},
  journal={The Journal of nervous and mental disease},
  volume={175},
  number={9},
  pages={526--536},
  year={1987},
  publisher={LWW}
}

@article{munby1989reflection,
  title={Reflection-in-action and reflection-on-action},
  author={Munby, Hugh},
  journal={Current issues in education},
  volume={9},
  number={1},
  pages={31--42},
  year={1989},
  publisher={Purdue University Press}
}

@inproceedings{sengers2005reflective,
  title={Reflective design},
  author={Sengers, Phoebe and Boehner, Kirsten and David, Shay and Kaye, Joseph'Jofish'},
  booktitle={Proceedings of the 4th decennial conference on Critical computing: between sense and sensibility},
  pages={49--58},
  year={2005}
}

@inproceedings{zhang2024s,
  title={“It's a Fair Game”, or Is It? Examining How Users Navigate Disclosure Risks and Benefits When Using LLM-Based Conversational Agents},
  author={Zhang, Zhiping and Jia, Michelle and Lee, Hao-Ping and Yao, Bingsheng and Das, Sauvik and Lerner, Ada and Wang, Dakuo and Li, Tianshi},
  booktitle={Proceedings of the 2024 CHI Conference on Human Factors in Computing Systems},
  pages={1--26},
  year={2024}
}

@inproceedings{vekaria2025big,
  title={Big Help or Big Brother? Auditing Tracking, Profiling, and Personalization in Generative $\{$AI$\}$ Assistants},
  author={Vekaria, Yash and Canino, Aurelio Loris and Levitsky, Jonathan and Ciechonski, Alex and Callejo, Patricia and Mandalari, Anna Maria and Shafiq, Zubair},
  booktitle={34th USENIX Security Symposium (USENIX Security 25)},
  pages={8115--8134},
  year={2025}
}

@article{nissenbaum2004privacy,
  title={Privacy as contextual integrity},
  author={Nissenbaum, Helen},
  journal={Wash. L. Rev.},
  volume={79},
  pages={119},
  year={2004},
  publisher={HeinOnline}
}

@article{fang2026ai,
  title={AI-Wrapped: Participatory, Privacy-Preserving Measurement of Longitudinal LLM Use In-the-Wild},
  author={Fang, Cathy Mengying and Karny, Sheer and Archiwaranguprok, Chayapatr and Samaradivakara, Yasith and Pataranutaporn, Pat and Maes, Pattie},
  journal={arXiv preprint arXiv:2602.18415},
  year={2026}
}

@inproceedings{hill1992editwear,
  title={Edit Wear and Read Wear},
  author={Hill, Will and Hollan, Jim and Wroblewski, Dave and McCandless, Tim},
  booktitle={Proceedings of the SIGCHI Conference on Human Factors in Computing Systems},
  pages={3--9},
  year={1992},
  doi={10.1145/142750.142751}
}

@inproceedings{erickson1999socially,
  title={Socially Translucent Systems: Social Proxies, Persistent Conversation, and the Design of {Babble}},
  author={Erickson, Thomas and Smith, David N. and Kellogg, Wendy A. and Laff, Mark and Richards, John T. and Bradner, Erin},
  booktitle={Proceedings of the SIGCHI Conference on Human Factors in Computing Systems},
  pages={72--79},
  year={1999},
  doi={10.1145/302979.303017}
}

@article{benk2026same,
  title={Same Performance, Hidden Bias: Evaluating Hypothesis-and Recommendation-Driven AI},
  author={Benk, Michaela and Miller, Tim},
  journal={arXiv preprint arXiv:2603.15824},
  year={2026}
}

@article{erickson2000social,
  title={Social Translucence: An Approach to Designing Systems that Support Social Processes},
  author={Erickson, Thomas and Kellogg, Wendy A.},
  journal={ACM Transactions on Computer-Human Interaction},
  volume={7},
  number={1},
  pages={59--83},
  year={2000},
  doi={10.1145/344949.345004}
}

@inproceedings{wei2016grouplink,
  title={GroupLink: Group Event Recommendations Using Personal Digital Traces},
  author={Wei, Jialin and Chandramouli, Badrish and Subramanian, Lakshminarayanan and Wu, Denny and Li, Tiancheng and Qian, Xinyu and Sezgin, Metin and Ji, Yong and Gao, Jianfeng and Acquisti, Alessandro},
  booktitle={Proceedings of the 19th ACM Conference on Computer-Supported Cooperative Work \& Social Computing Companion},
  pages={149--152},
  year={2016},
  doi={10.1145/2818052.2869126}
}

@inproceedings{chang2021tab,
  title={When the Tab Comes Due: Challenges in the Cost Structure of Browser Tab Management},
  author={Chang, Joseph Chee and Kittur, Aniket and Hahn, Nathan and Myers, Brad A.},
  booktitle={Proceedings of the 2021 CHI Conference on Human Factors in Computing Systems},
  pages={1--13},
  year={2021},
  doi={10.1145/3411764.3445585}
}

@inproceedings{leshed2008coscripter,
  title={CoScripter: Automating \& Sharing How-To Knowledge in the Enterprise},
  author={Leshed, Gilly and Haber, Eben and Matthews, Tara and Lau, Tessa},
  booktitle={Proceedings of the SIGCHI Conference on Human Factors in Computing Systems},
  pages={1719--1728},
  year={2008},
  doi={10.1145/1357054.1357323}
}

@article{zhao2024wildchat,
  title={WildChat: 1M ChatGPT Interaction Logs in the Wild},
  author={Zhao, Wenting and Ren, Xiang and Hessel, Jack and Cardie, Claire and Choi, Yejin and Deng, Yuntian},
  journal={arXiv preprint arXiv:2405.01470},
  year={2024}
}

@article{niu2026ethical,
  title={A Literature Review of Ethical Considerations in Recommender Systems for User-Generated Content in Human-Computer Interaction},
  author={Niu, Shuo and Li, Tianyi and Chi, Mohan},
  journal={ACM Transactions on Recommender Systems},
  volume={4},
  number={2},
  year={2026},
  doi={10.1145/3770747}
}

@article{tao2026care,
  title={An LLM Chatbot to Facilitate Primary-to-Specialist Care Transitions: A Randomized Controlled Trial},
  author={Tao, Xinge and Zhou, Shuya and Ding, Kai and Li, Sairan and Li, Yanzeng and Wu, Boyou and Huang, Qirui and Chen, Wangyue and Shen, Muzi and Meng, En and others},
  journal={Nature Medicine},
  year={2026},
  doi={10.1038/s41591-025-04176-7}
}

@inproceedings{zhang2024dismouse,
  title={DisMouse: Disentangling Information from Mouse Movement Data},
  author={Zhang, Guanhua and Hu, Zhiming and Bulling, Andreas},
  booktitle={Proceedings of the 37th Annual ACM Symposium on User Interface Software and Technology},
  year={2024},
  doi={10.1145/3654777.3676411}
}

@misc{yang2026lessonsrealworlddeploymentcognitionpreserving,
  title={Lessons from Real-World Deployment of a Cognition-Preserving Writing Tool: Students Actively Engage with Critical Thinking and Planning Affordances},
  author={Yang, Yinuo and Zhang, Zheng and Tang, Ningzhi and Wang, Xu and Ambrose, Alex and Myers, Nathaniel and Clauss, Patrick and Li, Toby Jia-Jun},
  year={2026},
  eprint={2603.15777},
  archivePrefix={arXiv},
  primaryClass={cs.HC},
  url={https://arxiv.org/abs/2603.15777}
}

@inproceedings{peng2025glitter,
  title={Glitter: An AI-Assisted Platform for Material-Grounded Asynchronous Discussion in Flipped Learning},
  author={Peng, Weirui and Yang, Yinuo and Zhang, Zheng and Li, Toby Jia-Jun},
  booktitle={Proceedings of the 38th Annual ACM Symposium on User Interface Software and Technology},
  pages={1--22},
  year={2025}
}

@inproceedings{yang2025spark,
  title={SPARK: Real-Time Monitoring of Multi-Faceted Programming Exercises},
  author={Yang, Yinuo and Zhang, Ashley Ge and Oney, Steve and Wang, April Yi},
  booktitle={2025 IEEE Symposium on Visual Languages and Human-Centric Computing (VL/HCC)},
  pages={81--92},
  year={2025},
  organization={IEEE}
}

@article{zhang2026editrail,
  title={Editrail: Understanding AI Usage by Visualizing Student-AI Interaction in Code},
  author={Zhang, Ashley Ge and Jhou, Yan-Ru and Yang, Yinuo and Rao, Shamita and Arab, Maryam and Chen, Yan and Oney, Steve},
  journal={arXiv preprint arXiv:2601.20085},
  year={2026}
}

@inproceedings{yang2024vizcode,
  title={Vizcode: A practical real-time tool for in-class computer programming tutoring},
  author={Yang, Yinuo and Oney, Steve},
  booktitle={Proceedings of the Eleventh ACM Conference on Learning@ Scale},
  pages={544--546},
  year={2024}
}

@article{gao2023cirs,
  title={CIRS: Bursting filter bubbles by counterfactual interactive recommender system},
  author={Gao, Chongming and Wang, Shiqi and Li, Shijun and Chen, Jiawei and He, Xiangnan and Lei, Wenqiang and Li, Biao and Zhang, Yuan and Jiang, Peng},
  journal={ACM Transactions on Information Systems},
  volume={42},
  number={1},
  pages={1--27},
  year={2023},
  publisher={ACM New York, NY, USA}
}

@article{loewenstein1992anomalies,
  title={Anomalies in intertemporal choice: Evidence and an interpretation},
  author={Loewenstein, George and Prelec, Drazen},
  journal={The quarterly journal of economics},
  volume={107},
  number={2},
  pages={573--597},
  year={1992},
  publisher={MIT Press}
}

@misc{everydayconversations2024,
  author = {Hugging Face},
  title = {Everyday Conversations for LLMs},
  year = {2024},
  howpublished = {\url{https://huggingface.co/datasets/HuggingFaceTB/everyday-conversations-llama3.1-2k}}
}

@article{grimm2010social,
  title={Social desirability bias},
  author={Grimm, Pamela},
  journal={Wiley international encyclopedia of marketing},
  year={2010},
  publisher={Wiley Online Library}
}

@inproceedings{shen2025bidirectional,
  title={Bidirectional Human-AI Alignment: Emerging Challenges and Opportunities},
  author={Shen, Hua and Knearem, Tiffany and Ghosh, Reshmi and Liu, Michael Xieyang and Monroy-Hern{\'a}ndez, Andr{\'e}s and Wu, Tongshuang and Yang, Diyi and Huang, Yun and Mitra, Tanushree and Li, Yang and others},
  booktitle={Proceedings of the Extended Abstracts of the CHI Conference on Human Factors in Computing Systems},
  pages={1--6},
  year={2025}
}

@article{shani2024multi,
  title={Multi-turn reinforcement learning with preference human feedback},
  author={Shani, Lior and Rosenberg, Aviv and Cassel, Asaf and Lang, Oran and Calandriello, Daniele and Zipori, Avital and Noga, Hila and Keller, Orgad and Piot, Bilal and Szpektor, Idan and others},
  journal={Advances in Neural Information Processing Systems},
  volume={37},
  pages={118953--118993},
  year={2024}
}

@misc{chen2026digitalcoach,
      title={DigitalCoach: Communication and Grounding Gaps in Human and Agentic Computer Use Coaching}, 
      author={Meng Chen and Anya Ji and Tsung-Han Wu and Tobias Maringgele and David M. Chan and Alane Suhr and Amy Pavel},
      year={2026},
      eprint={2606.31980},
      archivePrefix={arXiv},
      primaryClass={cs.CL},
      url={https://arxiv.org/abs/2606.31980}, 
}

@misc{chen2026taskartisan,
      title={TaskArtisan: Designing Composable Generative Widgets for LLM-Assisted Analysis}, 
      author={Meng Chen and Amy Pavel},
      year={2026},
      eprint={2607.17394},
      archivePrefix={arXiv},
      primaryClass={cs.HC},
      url={https://arxiv.org/abs/2607.17394}, 
}

@book{dewey2022we,
  title={How we think},
  author={Dewey, John},
  year={2022},
  publisher={DigiCat}
}

% \newpage
% \appendix
\clearpage

\onecolumn

\appendix

\section{Additional Extension Screenshots}

\begin{figure}[h]
    \centering
    \includegraphics[width=0.8\linewidth]{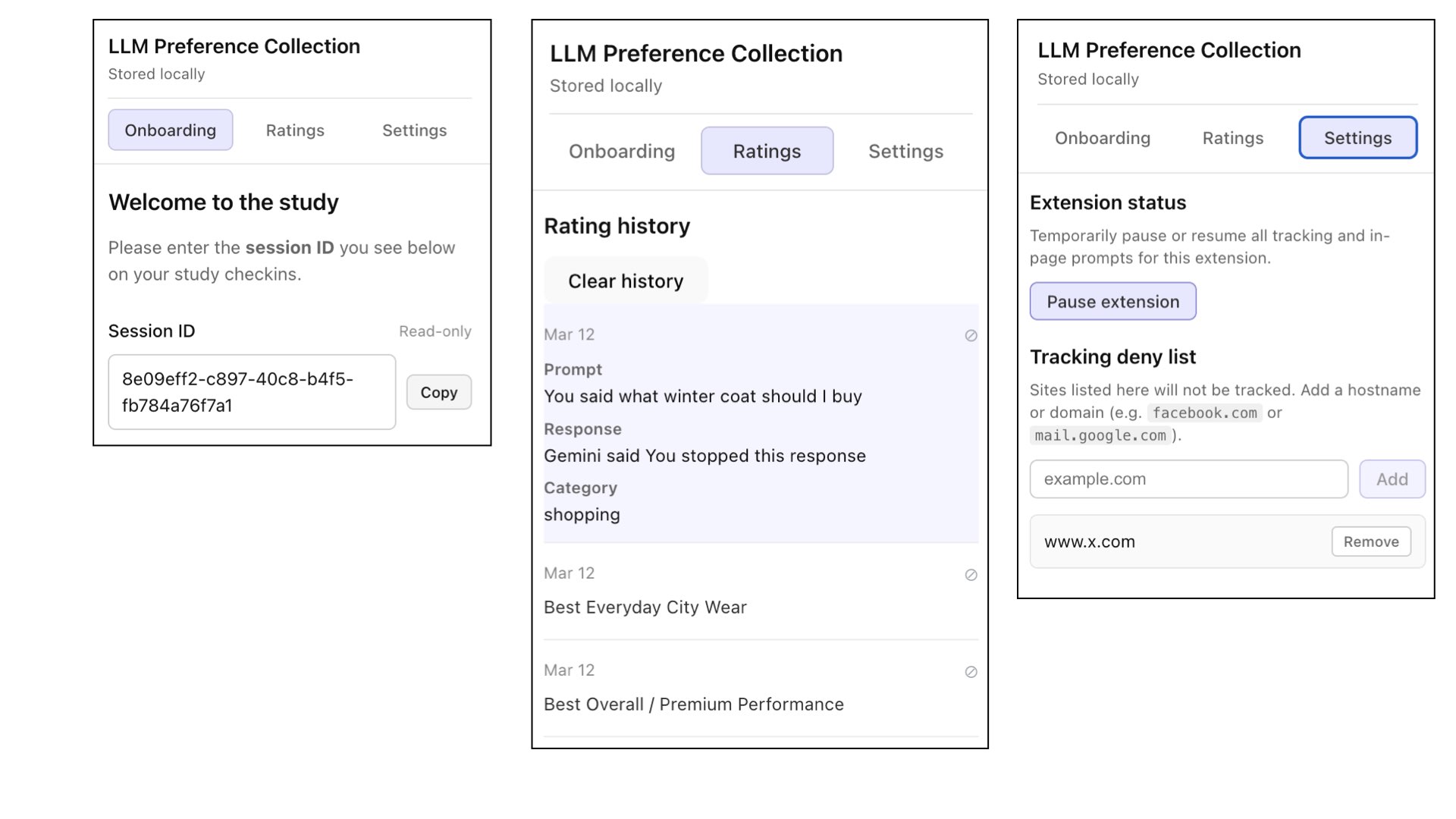}
    \caption{Screenshots of the Chrome extension used for longitudinal alignment data collection. (A) Onboarding interface where users register their session ID. (B) Ratings history view showing captured LLM prompts, responses, and inferred categories. (C) Settings panel providing user controls for pausing data collection and specifying websites to exclude from tracking.}
    \Description{Three side-by-side screenshots of the Chrome extension interface. The first panel shows the onboarding screen with a session ID field. The second panel displays the ratings history view with previously captured prompts, responses, and inferred topic categories. The third panel shows the settings screen with controls for pausing the extension and specifying websites to exclude from tracking.}
    \label{fig:ext_screenshots}
\end{figure}

\section{Topic Classification Prompt}
\label{app:topic_classification_prompt}

\begin{mybox}{Prompt used for LLM-based Topic Classification}

You classify the conversation into exactly ONE category. Choose the single best match. If none match return other. 

Follow the following definitions.

shopping: product recommendations, what to buy, prices, deals, gifts, reviews, clothing/footwear/fashion

job\_career: job applications, resume/cover letter help, career advice, grad school applications, interviews

travel: trip planning, destinations, conversation about flights, hotels, itineraries, travel recommendations

homework: homework help, assignment support, tutoring, studying, exam prep

email\_drafting: drafting emails, composing messages, professional correspondence

relationship: personal relationship advice, dating, family, friendship, interpersonal issues

restaurant: restaurant recommendations, dining suggestions, reservations, food spots

fitness: exercise, workout plans, nutrition for fitness, health routines

productivity: time management, organization, task management, productivity tips

other: none of the above (coding, creative writing, general knowledge, etc.)

Respond with exactly two lines:

Line 1: The category (one word, lowercase, e.g. shopping, job\_career, travel, homework, email\_drafting, 

relationship, restaurant, fitness, productivity, or other)

Line 2: A brief one sentence reason.

\end{mybox}

\section{Technical Evaluation}
\label{app:technical_evaluation}

% \hl{We measured how often the domain classifier successfully detected interactions within tracked categories and how frequently follow-up events were triggered.}

We conducted a technical evaluation of the domain classification prompt used to detect whether a user’s interaction belongs to one of the tracked categories supported by our system. To assess classification performance, we benchmarked the prompt on the public dataset \textit{everyday-conversations-llama3.1-2k}~\cite{everydayconversations2024}. The dataset contains 2380 simple multi-turn conversations spanning diverse everyday topics and subtopics. We sampled 200 data points for this evaluation.

We then ran the exact same classification prompt described in Section~\ref{app:topic_classification_prompt} using a locally hosted Llama 3 model and an API call to OpenAI GPT models. For each conversation, the model was given the user query and required to output exactly one category label. Accuracy was computed as the proportion of conversations for which the predicted category matched the mapped ground-truth label.
Table~\ref{tab:model_performance} reports the topic classification accuracy across the three evaluated models. Overall, all models achieved reasonably strong performance on the held-out evaluation set, with accuracies ranging from 0.77 to 0.84. \texttt{GPT-4o-mini} achieved the highest accuracy (0.84), followed by \texttt{GPT-5} (0.81) and the locally hosted \texttt{Llama 3.1 70B} model (0.77).  Overall, the results indicate that the prompt produced stable classification performance across different model backends.

\begin{table}[]
\begin{tabular}{@{}lll@{}}
\toprule
Model & Version & Accuracy \\ \midrule
GPT-4o & gpt-4o-mini-2024-07-18 & .84 \\
GPT-5 & gpt-5-2025-08-07 & .81 \\
Llama & llama3.1:70b &  .77 \\ \bottomrule
\end{tabular}
\caption{Topic classification performance across different models}
\label{tab:model_performance}
\Description{Table reporting topic classification accuracy across three evaluated models. GPT-4o-mini achieves the highest accuracy at 0.84, followed by GPT-5 at 0.81, and Llama 3.1 70B at 0.77. The table shows that all three models perform reasonably well.}
\end{table}

\section{Post-Interaction Response Evaluation Questionnaire}
\label{app:response_eval_questions}

Participants were asked to rate each response using the following items on a 5-point Likert scale 
(\textit{1 = Not at all, 5 = Extremely}).

\begin{enumerate}
    \item Was the response helpful?
    \item How accurate did you find the information provided in the response?
    \item How relevant was the response to your specific needs?
    \item To what extent do you trust the advice or data given?
    \item Was the response easy to understand and follow?
    \item How harmful was the response?
\end{enumerate}

\end{document}